\documentclass{article}
\usepackage[dvips]{epsfig}
  \title{The masses of the mesons and baryons \\
  Part  II.  The standing wave model}
  \author{E.~L. Koschmieder and T.~H. Koschmieder \\
   {\small Center for Statistical Mechanics,  The University of Texas at
   Austin,} \\ {\small Austin,  TX 78712,  USA}}
  \date{\today}

\begin{document}    
\twocolumn[\maketitle
\begin{center} \parbox{4.3in}{\small
In order to explain the empirical integer multiple rule for the stable
mesons and baryons presented in the preceding paper we assume that the
particles are held together in a cubic nuclear lattice.  This is a novel
approach to the particles, based on the fact that the range of the weak
nuclear force is only a thousandth of the diameter of the nucleon, and
that the crystals are the best-known macroscopic bodies held together by
a microscopic force.  We investigate the standing waves in a cubic
nuclear lattice.  From the frequency distribution of the waves follows
that the masses of the $\gamma$-branch particles are integer multiples
of $m(\pi^0)$.  We show that each particle has automatically an
antiparticle.  Assuming that the energy of the oscillations is
determined by Planck's formula for the energy of a linear oscillator, it
turns out that the $\pi^0$ meson and the other members of the
$\gamma$-branch are like cubic black bodies filled with plane, standing
electromagnetic waves.  Our standing wave model explains the integer
multiple rule of the masses of the neutral mesons and baryons of the
$\gamma$-branch and uses nothing else but photons.  Our results justify
the cubic lattice assumption. \\[1ex]
} \end{center}  \vspace*{10ex}]

\section{Introduction}

The masses of the so-called stable elementary particles are the
best-known and most characteristic property of the particles, but have
not yet been explained.  In a preceding article [1] we have shown that
it follows from the well-known decays of the elementary particles that
the spectrum of the particles splits into a $\gamma$-branch and a
neutrino branch.  From the masses of the particles follows that the
masses of the $\gamma$-branch particles are, within 3\%, integer
multiples of the $\pi^0$ meson and that the masses of the $\nu$-branch
are integer multiples of the $\pi^\pm$ mesons times a factor $0.86\pm
0.02$.  For the masses and decays see Table I and II of [1].  The
integer multiple rule suggests that the particles are the solution of a
wave equation.  In a previous article [2], we have tried to explain the
masses of the elementary particles by monochromatic eigenfrequencies of
standing waves in an elastic cube.  These eigenfrequencies depend on the
value of Young's modulus of the material of the cube.  In a subsequent
paper [3], we have explained the value of Young's modulus of the
material of the elementary particles with the help of Born's classical
model of cubic crystals, assuming that the crystal is held together by
the weak nuclear force acting between the lattice points.  The
explanation of the particles by monochromatic eigenfrequencies does not
seem to be tenable because a monochromatic frequency does not
accommodate the multitude of frequencies created in a high energy
collision of $10^{-23}$ sec duration.  We will now study whether the
so-called stable elementary particles of the $\gamma$-branch cannot be
described by the frequency spectrum of standing waves in a lattice, 
which can accommodate automatically the Fourier frequency spectrum of an
extreme short-time collision.

The investigation of the consequences of lattices for particle theory
was initiated by Wilson [4] who studied a fermion lattice.  This study
has developed over time into lattice QCD.  The results of such endeavors
have culminated in the paper of Weingarten [5] and his colleagues which
required elaborate year long numerical calculations.  They determined
the masses of seven particles ($K^\ast $, $p$, $\phi$, $\Delta$,
$\Sigma$, $\Xi$, $\Omega$), with an uncertainty of up to $\pm 8$\%,
agreeing with the observed particles within a few percent, up to 6\%.
Our theory covers all particles of the $\gamma$-branch, namely the
$\pi^0$, $\eta$, $\eta^\prime$, $\Lambda$, $\Sigma^0$, $\Xi^0$,
$\Omega^-$, $\Lambda_c^+$, $\Sigma_c^0$, $\Xi_c^0$, and $\Omega_c^0$
particles and agrees with the measured masses within at most 3.3\%.  The
masses of the $\nu$-branch can be explained separately by standing waves
in a neutrino lattice.

\section{The frequency spectrum of the \\
 oscillations of a cubic  lattice}

It will be necessary to outline the most elementary aspects of the
theory of lattice oscillations since we will, in the following,
investigate whether the masses of the elementary particles can be
explained with the help of the frequency spectra of standing waves in a
cubic lattice.  The classic paper describing lattice oscillations is
from Born and v.\ Karman [6], henceforth referred to as B\&K.  They
looked at first at the oscillations of a one-dimensional chain of points
with mass $m$, separated by a constant distance $a$.  B\&K assume that
the forces exerted on each point of the chain originate only from the
two neighboring points.  These forces are opposed to and proportional to
the displacements, as with elastic springs.  The equation of motion is
in this case
\begin{equation} \label{e1}
m\ddot{u}_n = \alpha \left( u_{n+1} - u_n \right) - \alpha \left( u_n
    - u_{n-1} \right) \; .
\end{equation}
The displacements of the mass points from their equilibrium position are
given by $u_n$.  The dots signify, as usual, differentiation with
respect to time, $\alpha$ is a constant characterizing the force between
the lattice points, and $n$ is an integer number.

In order to solve (\ref{e1}) B\&K set
\begin{equation} \label{e2}
u_n = A \mathrm{e}^{i (\omega t + n\phi)} \; , 
\end{equation}
which is obviously a temporally and spatially periodic solution. $n$
is an integer, with $n\le N$, where $N$ is the number of points in the
chain. At $2n\phi =\pi$ there are nodes, where for all times $t$ the
displacements are zero, as with standing waves. If a displacement
is repeated after $n$ points we have $na =\lambda$, where $\lambda$ is 
the wavelength, and it must be $n\phi =2\pi$ according to (\ref{e2}). It 
follows that $\lambda =2\pi a/\phi$. Inserting (\ref{e2}) into 
(\ref{e1}) one obtains a continuous frequency spectrum given by
\begin{equation} \label{e3}
\nu  = 2 \sqrt{\frac{\alpha}{m}} \; \sin (\phi /2) \; .
\end{equation}
B\&K point out that there is not only a continuum of frequencies, but
also a maximal frequency which is reached when $\phi = \pi$, or at the
minimum of the possible wavelengths $\lambda = 2a$.

B\&K then discuss the three-dimensional lattice, with lattice constant
$a$ and masses $m$, the monatomic case, i.e.\ when all lattice points
are of the same mass.  They reduce the complexities of the problem by
considering only the 18 points nearest to any point.  These are 6 points
at distance $a$, and 12 points at distance $a \sqrt{2}$.  B\&K assume
that the forces between the points are linear functions of the small
displacements, that the symmetry of the lattice is maintained, and that
the equations of motion transform into the equations of motion of
continuum mechanics for $a\rightarrow 0$.  We cannot reproduce the
lengthy equations of motion of the three-dimensional lattice.  In the
three-dimensional case we deal with the forces caused by the 6 points at
the distance $a$ which are characterized by the constant $\alpha$ in the
case of central forces.  There are also the forces which originate from
the 12 points at distance $a \sqrt{2}$, characterized by the constant
$\gamma$, which is important later on.  We investigate the propagation
of plane waves in a three-dimensional monatomic lattice with the ansatz
\begin{equation} \label{e4}
u_{\ell ,m} = u_0 \; \mathrm{e}^{i(\omega t +\ell \phi_1 +m \phi_2)}
\end{equation}
and a similar ansatz for $v_{\ell ,m}$, with $\ell ,m$ being integer
numbers $\le N$, where $N$ is the number of lattice points along a side 
of the cube.  We also consider higher order solutions, with $i_1 \cdot
\ell$ and $i_2\cdot m$, where $i_1 ,i_2$ are integer numbers.  The
boundary conditions are periodic.  The number of normal modes must be
equal to the number of particles in the lattice.  B\&K arrive, in the
case of two-dimensional waves, at a secular equation for the
frequencies
\begin{equation} \label{e5}
\left| \begin{array}{cc}
   A(\phi_1 ,\phi_2) - m\nu^2 & B(\phi_2 ,\phi_1) \\
   B(\phi_1 ,\phi_2) & A(\phi_2 ,\phi_1 ) - m\nu^2 
\end{array} \right| = 0 \; ,
\end{equation}
The formulas for $A(\phi_1 ,\phi_2)$ and $B(\phi_1 ,\phi_2)$ are given 
in equation (17) of B\&K.

The theory of lattice oscillations has been pursued in particular by
Blackman [7], a summary of his and other studies is in [8].
Comprehensive reviews of the results of linear studies of lattice
dynamics have been written by Born and Huang [9], and by Maradudin et
al.\ [10].

\section{The masses of the particles of the $\gamma$-branch}

We will now assume, as seems to be quite natural, that the particles of
the $\gamma$-branch \textit{consist of the same particles into which
they decay}, i.e., of photons.  We base this assumption on the fact that
photons are the principal mode of decay of the $\gamma$-branch
particles, the characteristic example is $\pi^0 \rightarrow \gamma
\gamma$ (98.8\%).  The composition of the particles of the
$\gamma$-branch suggested here offers a direct route from the formation
of a $\gamma$-branch particle, through its lifetime, to its decay
products.  Particles that are made of photons are necessarily neutral,
as the majority of the particles of the $\gamma$-branch are.

We also base our assumption that the particles of the $\gamma$-branch
are made of photons on the circumstances of the formation of the
$\gamma$-branch particles.  The most simple and straightforward creation
of a $\gamma$-branch particle is the reaction $\gamma + p \rightarrow
\pi^0 + p +\gamma^\prime$.  A photon impinges on a proton and creates a
$\pi^0$ meson.  In a timespan of order of $10^{-23}$ sec the pulse of
the incoming electromagnetic wave is, according to Fourier analysis,
converted into a continuum of electromagnetic waves with frequencies
ranging from $10^{23}$ sec$^{-1}$ to $\nu \rightarrow \infty$.  The wave
packet so created decays, according to experience, after $8.4\cdot
10^{-17}$ sec, into two electromagnetic waves or $\gamma$-rays.  It
seems to be very unlikely that Fourier analysis does not hold for the
case of an electromagnetic wave impinging on a proton.  The question
then arises of what happens to the electromagnetic waves in the timespan
of $10^{-16}$ seconds between the creation of the wave packet and its
decay into two $\gamma$-rays?  We will investigate whether the
electromagnetic waves cannot continue to exist for the $10^{-16}$
seconds until the wave packet decays.

There must be a mechanism which holds the wave packet of the newly
created particle together, or else it will disperse.  We assume that the
very many photons in the new particle are held together in a cubic
lattice.  If a particle consists of photons, alternate photons must have
opposite spin, otherwise the spin of the particle could not be zero.
Ordinary cubic lattices, such as the NaCl lattice, are held together by
attractive forces between particles of opposite polarity.  We assume
that the photon lattice is held together by weak attractive forces
between photons of opposite polarity.  Electrodynamics does not predict
the existence of such a force between two photons.  However, electroweak
theory says that $e^2 \approx g^2$, and we will now assume that there is
a corresponding force in the photon lattice.  It is not unprecedended
that photons have been considered to be building blocks of the
elementary particles.  Schwinger [11] has once studied an exact
one-dimensional quantum electrodynamical model in which the photon
acquired a mass $\sim e^2$.

We will now investigate the standing waves of a cubic photon lattice.
We assume that the lattice is held together by a weak force acting from
one lattice point to the next.  We assume that the range of this force
is $10^{-16}$ cm, because the range of the weak nuclear force is of
order of $10^{-16}$ cm, according to [12].  Likewise, we assume that the
sidelength of the lattice is about $10^{-13}$ cm, which follows from the
size of the nucleon as given by [13].  There are then $10^9$ lattice
points.  Because it is the most simple case, we assume that a central
force acts between lattice points of different polarity.  We cannot
consider spin, isospin, strangeness or charm.  The frequency equation
for the two-dimensional oscillations of an isotropic monatomic cubic
lattice with central forces is
\begin{equation} \label{e6}
  \nu^2 = \frac{2\alpha}{4\pi^2 M} \; \left\{ 2 -\cos \phi_1 \,
  \cos \phi_2 + \sin \phi_1 \sin \phi_2 - \cos \phi_1 \right\} \, .
\end{equation}
In the isotropic case, i.e.\ when $\gamma /\alpha =0.5$, Eq.\ (\ref{e6})
follows directly from the equation of motion for the displacements in a
monatomic lattice, which are given e.g.\ by Blackman [7].  The minus
sign in front of $\cos \phi_1$ means that the waves are longitudinal.
All frequencies that solve (\ref{e6}) come with either a plus or a minus
sign, which is, as we will see, important.

\begin{figure}[tb!] \begin{center}
      \epsfig{file=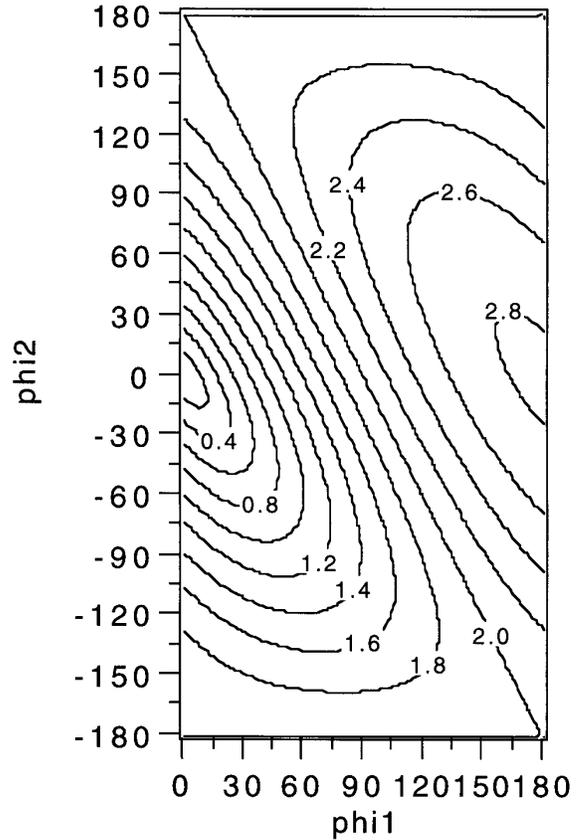,width=3in}
\end{center}
\caption{The frequency distribution $\nu_- /\nu_0$ of the basic mode 
according to Eq.\ (\ref{e6}). Monatomic, isotropic case. Resolution of 
calculation: 1 degree.} \label{f1} \end{figure}

The frequency distribution following from (\ref{e6}) is shown in Fig.\
\ref{f1}. There are some easily verifiable frequencies.  For example at
$\phi_1, \phi_2 =0,0$ it is $\nu =0$, at $\phi_1 , \phi_2=\pi /2$, $\pi
/2$ it is $\nu = \nu_0 \sqrt{6}$, at $\phi_1 , \phi_2 = \pi / 2, -\pi /
2$ we have $\nu = \nu_0 \sqrt{2}$.  Furthermore at $\phi_1 , \phi_2 =
\pi , 0$ we have $\nu =\nu_0 \sqrt{8}$, and for all values of $\phi_1$
it is $\nu =2\nu_0$ at $\phi_2 = \pi$ and $\phi_2 =-\pi$, with $\nu_0
=\sqrt{\alpha /4\pi^2 M}$, or as we will see $\nu_0 = c_* /2\pi a$.

The limitation of the group velocity in the photon lattice has now to be
considered.  The formula for the group velocity is given by
\begin{equation} \label{e7}
c_g = \frac{d\omega}{dk} = a \sqrt{\frac{\alpha}{M}} \cdot \frac{df( 
\phi_1 , \phi_2 )}{d\phi} \; .
\end{equation}
The group velocity in the photon lattice has to be equal to the velocity
of light $c_*$, throughout the entire frequency spectrum, because
photons move with the velocity of light.  In order to learn how this
requirement affects the frequency distribution we have to know the value
of $\sqrt{\alpha /M}$ in a photon lattice.  But we do not have
information about what either $\alpha$ or $M$ might be in this case.  We
assume in the following that $a\sqrt{\alpha /M} = c_*$, which means,
since $a = 10^{-16}$ cm, that $\sqrt{\alpha /M} = 3\cdot 10^{26}$
sec$^{-1}$, or that the corresponding period is $\tau = \frac{1}{3}\cdot
10^{-26}$ sec, which is the time it takes for a wave to travel with the
velocity of light over one lattice distance.  With $a\sqrt{\alpha /M} =
c_*$, the equation for the group velocity is
\begin{equation} \label{e8}
c_g = c_* \cdot \frac{df}{d\phi} \; .
\end{equation}

\begin{figure}[tb!] \begin{center}
      \epsfig{file=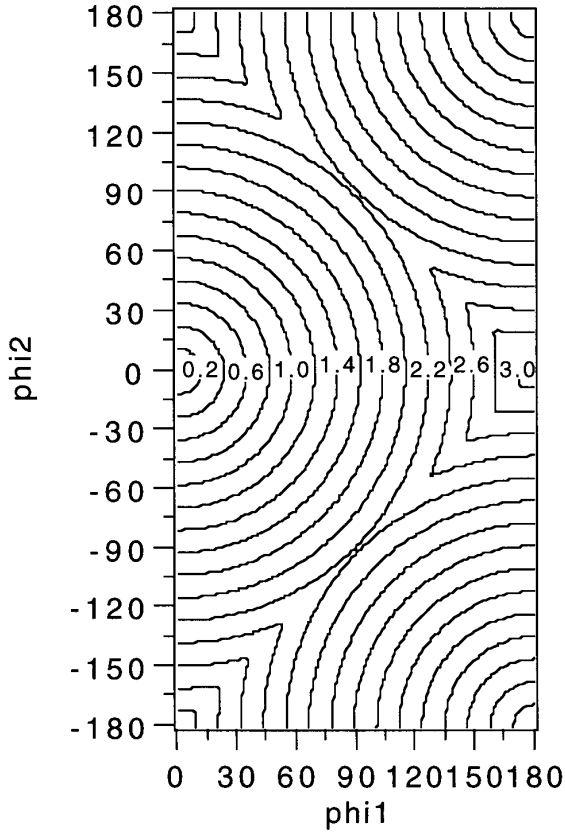,width=3in}
\end{center}
\caption{The corrected frequency distribution $\nu_- /\nu_0$ of the 
basic mode. Monatomic, isotropic case. Resolution of calculations: 1 
degree.} \label{f2} \end{figure}

For a photon lattice that means, since $c_g$ must then always be equal
to $c_*$, that $df / d\phi =1$.  This requirement determines the form of
the frequency distribution, regardless of the order of the mode of
oscillation.  The frequencies of the corrected spectrum must increase
from the origin $\phi_1 ,\phi_2 = 0, 0$ with slope 1 until a maximum is
reached, from where the frequency must decrease with slope 1 to $\nu=0$.
The frequency distribution corrected for (\ref{e8}) is shown in Fig.\ 
\ref{f2}. The corrected frequency distributions of higher modes are of 
the
same type, but for the area they cover.  The second mode ($i_1, i_2 =2$)
covers 4 times the area of the basic mode, because $2\phi$ ranges from 0
to $2\pi$, (for $\phi >0$), whereas the basic mode ranges from 0 to
$\pi$.  Consequently the energy $E$ (Eq.\ \ref{e9}) contained in all
oscillations of the second mode is four times larger than the energy of
the basic mode, because the energy contained in the lattice oscillations
must be proportional to the sum of all frequencies.  Adding, by
superposition, to the second mode different numbers of basic modes or
second modes will give exact integer multiples of the energy of the
basic mode.  Now we understand the integer multiple rule of the
particles of the $\gamma$-branch.  There is, in the framework of this
theory, on account of Eq. (\ref{e8}), no alternative but integer 
multiples
of the basic mode for the energy contained in the oscillations of the
different modes or for superpositions of different modes.  In other
words, the masses of the different particles are integer multiples of
the mass of the $\pi^0$ meson, assuming that there is no spin, isospin,
strangeness or charm.  We remember that the measured masses in Table I
of [1], which incorporate different spins, isospins, strangeness, and
charm spell out the integer multiples rule within 3\%\ accuracy.  It is
worth noting that \textit{there is no free parameter} if one takes the
ratio of the energies contained in the frequency distributions of the
different modes, because the factor $\sqrt{\alpha /M}$ in Eq.\ 
(\ref{e6})
cancels.  This means, in particular, that the ratios of the frequency
distributions, or the mass ratios, are independent of the mass of the
photons at the lattice points, as well as of the magnitude of the force
between the lattice points.

Let us summarize our findings concerning the particles of the
$\gamma$-branch.  The $\pi^0$ meson is the basic mode of the photon
lattice oscillations.  The $\eta$ meson corresponds to the first higher
mode ($i_1 ,i_2=2$), as is suggested by $m(\eta )\approx 4m(\pi^0)$.
$\eta^\prime$ is a superposition of three basic modes on the first
higher mode.  The $\Lambda$ particle corresponds to the superposition of
two higher modes ($i_1 ,i_2 =2$), as is suggested by $m(\Lambda)\approx
2m(\eta )$.  This superposition apparently results in the creation of
spin $\frac{1}{2}$.  The two modes would then have to be coupled.  The
$\Sigma^0$ and $\Xi^0$ baryons are superpositions of one or two basic
modes on the $\Lambda$ particle.  The $\Omega^-$ particle corresponds to
the superposition of three coupled higher modes ($i_1 , i_2 = 2$) as is
suggested by $m(\Omega^- ) \approx 3m(\eta )$.  This procedure 
apparently causes
spin $\frac{3}{2}$.  The charmed $\Lambda_c^+$ particle seems to be the
first particle incorporating a (3.3) mode.  $\Sigma_c^0$ is apparently
the superposition of a basic mode on $\Lambda_c^+$, as is suggested by
the decay of $\Sigma_c^0$.  The easiest explanation of $\Xi_c^0$ is that
it is the superposition of two coupled (3.3) modes.  The superposition
of two modes of the same type is, as in the case of $\Lambda$,
accompanied by spin $\frac{1}{2}$.  The $\Omega_c^0$ baryon is
apparently the superposition of two basic modes on the $\Xi_c^0$
particle.  All neutral particles of the $\gamma$-branch are thus
accounted for, the agreement between the measured masses and the
theoretical values is $\le 3$\%, see Table I of [1].  We find it
interesting that all $\gamma$-branch particles with coupled $2\cdot
(2.2)$ modes, or the $\Omega^-$ particle with the coupled $3\cdot (2.2)$
mode, have strangeness.  But this rule does not hold in the presence of
a (3.3) mode.  All $\gamma$-branch particles with a (3.3) mode have
charm.  We have also found the $\gamma$-branch \textit{antiparticles},
which follow from the negative frequencies which solve Eq.\ (\ref{e6}).
Antiparticles have always been associated with negative energies.
Following Dirac's argument for electrons and positrons, we associate the
masses following from the negative frequency distributions with
antiparticles.  We emphasize that the existence of antiparticles is an
automatic consequence of our theory.

\section{The  mass  of  the $\pi^0$ meson}

So far we have studied the ratios of the masses of the particles.  We
will now determine the mass of the $\pi^0$ meson in order to validate
that the mass ratios connect with the actual masses of the particles.
The energy of the $\pi^0$ meson is $E(m(\pi^0))=134.976$ MeV =
$2.1626\cdot 10^{-4}$ erg.  For the energy $E$ of all oscillations we
use the equation
\begin{equation} \label{e9}
E = \frac{Nh\nu_0}{(2\pi)^2} \; \int \int_0^{2\pi} f(\phi_1 , \phi_2 
   ) \; d\phi_1 \; d\phi_2 \; , 
\end{equation}

This equation originates from Born and v.~Karman.  $N$ is the number of
lattice points.  For $f(\phi_1, \phi_2)$ we use our Eq.\ (\ref{e6}). 
Using the frequency distribution shown in Fig.\ \ref{f2} it turns out
that the numerical value of the double integral in (9) is 66.896
(radians$^2$) for the corrected (1.1) state.  With $N=10^9$ and
$\nu_0=3\cdot 10^{26} /2\pi$ it follows from Eq.\ (\ref{e9}) that
$E({\rm corr})(1.1)$ is $5.36\cdot 10^8$ erg, that means $2.48\cdot
10^{12}$ times larger than $E(m(\pi^0))$.

In order to eliminate this discrepancy we use, instead of the simple
form $E=h\nu$, the complete quantum mechanical energy of a linear
oscillator as given by Planck,
\begin{equation} \label{e10}
E=\frac{h\nu}{{\rm e}^\frac{h\nu}{kT} -1} \; .
\end{equation}
This equation was already used by B\&K for the determination of the
specific heat of cubic crystals or solids.  Equation (\ref{e10}) calls 
into question the value of the temperature $T$ in the interior of a
particle. We determine $T$ empirically with the formula for the internal
energy of solids
\begin{equation} \label{e11}
u=\frac{R\Theta}{{\rm e}^{\Theta/T} -1} \; ,
\end{equation}
which is from Sommerfeld [14].  In this equation it is now $R= 10^9 k$,
where $k$ is Boltzmann's constant, and $\Theta$ is the characteristic
temperature introduced by Debye [15] for the explanation of the specific
heat of solids.  It is $\Theta =h\nu_m/k$, where $\nu_m$ is a maximal
frequency.  In the case of the oscillations making up the $\pi^0$ meson
the maximal frequency is $\nu_m =\pi \nu_0$, see Figs.\ 
\ref{f1},\ref{f2}, therefore $\nu_m =1.5\cdot 10^{26}$ sec$^{-1}$, and
we find that $\Theta = 7.199\cdot 10^{15}$ K.

In order to determine $T$ we set the internal energy $u$ equal to
$m(\pi^0 )c^2_*$.  It then follows from (\ref{e11}) that $\Theta /T = 
29.16$, or $T= 2.47\cdot 10^{14}$ K. That means that Planck's formula
(\ref{e11}) introduces a factor $1/({\rm e}^{\Theta /T}-1) \approx
1/e^{29.16} = 1/4.613\cdot 10^{12}$ into Eq.\ (\ref{e9}).  In other 
words, if we determine the temperature $T$ of the particle empirically
through equation (\ref{e11}), then we arrive from (\ref{e9}) at a mass 
of the $\pi^0$ meson of $1.16\cdot
10^{-4}$ erg.  The difference between the exact $E(m(\pi^0)) =2.16\cdot
10^{-4}$ erg and our calculated $E$(corr)(1.1), which describes the
$\pi^0$ meson, is well within the uncertainty of the number of the
lattice points and the lattice distance we have used.  If we take the
value of the \textit{radius} of the nucleon given in [13] verbatim,
$r=0.8\cdot 10^{-13}$ cm, and calculate the number of lattice points in
a sphere of that radius, then there should be $2.14\cdot 10^9$ lattice
points in the particle.  Since $E$ is directly proportional to the
number of lattice points it follows that then $E$(corr)(1.1) =
1.15$\cdot E(m(\pi^0))$.  The energy in the mass of the $\pi^0$ meson
and the energy in the corresponding lattice oscillations agree very
well, considering the uncertainties of the parameters involved.  It can
be shown that the factor exp($\Theta /T$) remains constant for the
higher modes.

To summarize.  We find that the energy in the $\pi^0$ meson and the
other particles of the $\gamma$-branch are correctly given by the energy
of the standing waves, if the energy of the oscillations is determined
by Planck's formula for the energy of a linear oscillator.  The $\pi^0$
meson is like an adiabatic, cubic black body filled with standing
electromagnetic waves.  We know from Bose's work [16] that Planck's
formula applies to a photon gas as well.

\section{Conclusions}

Let us summarize what we have assumed and what we have learned from this
study.  In short, for each neutral meson and each neutral baryon of the
$\gamma$-branch we have found a simple mode of standing waves in a cubic
lattice, the ratio of the energies of which agree within 3\%
with the ratio of the energies of the masses of the particles.  In order
to arrive at this result, we have first assumed that the neutral mesons
and baryons consist of the same particles into which they decay, which
seems to be a quite natural assumption.  For the explanation of the
mesons and baryons of the $\gamma$-branch we use only photons, nothing
else.  We assume the existence of a weak force which holds the photon
lattice of the $\gamma$-branch together.  Then we apply the results of
the classical study of Born and v.~Karman, and subsequent studies, about
lattice oscillations to the particle lattice.  We determine the
frequency distributions and the energy contained in plane standing waves
in a cubic lattice.  The $\gamma$-rays in the photon lattice must move
with the velocity of light, and we impose the condition that the group
velocity is equal to the velocity of light.  From the frequency
distributions of the standing waves in the lattice follow the ratios of
the masses of the particles.

The masses of the $\gamma$-branch, the $\pi^0$, $\eta$, $\eta^\prime$,
$\Lambda$, $\Sigma^0$, $\Xi^0$, $\Omega^-$, $\Lambda_c^+$, $\Sigma_c^0$,
$\Xi_c^0$, and $\Omega_c^0$ particles are found to be integer multiples
of the mass of the $\pi^0$ meson, in agreement with what the data on the
particle masses strongly suggest.  The integer multiple rule is a
consequence of the standing wave structure of the particles.  It is
important to note that in our theory the ratios of the masses of the
$\gamma$-branch particles to the mass of the $\pi^0$ meson \textit{do
not depend} on the sidelength of the lattice, and the distance between
the lattice points, neither do they depend on the strength of the force
between the lattice points nor on the mass of the lattice points.  The
mass ratios are determined only by the spectra of the frequencies of the
standing waves in the lattice.  Since the equation determining the
frequency of the standing waves is quadratic it follows automatically
that for each positive frequency there is also a negative frequency of
the same absolute value, that means that for each particle there exists
also an \textit{antiparticle}.

We have then determined the mass of the $\pi^0$ meson which follows from
our theory.  This requires consideration of the number $N^3$ of the
lattice points and of the value of $\nu_0 =c_* /2\pi a$.  Assuming that
the energy of the oscillations is determined by Planck's formula for the
energy of a linear oscillator, we arrive at a mass of the $\pi^0$ meson
which differs from the experimentally determined $m(\pi^0)$ by 15\%,
which is well within the uncertainties of $N^3$ and $a$.  The $\pi^0$
meson is like a cubic black body filled with plane, standing
electromagnetic waves, whose wavelengths are integer multiples of the
lattice constant $a$.  A rather conservative explanation of the $\pi^0$
meson, and the $\gamma$-branch particles.  It is worth noting that in
the $\gamma$-branch of our model there is a continuous line leading from
the creation of a particle out of photons or electromagnetic waves
through the lifetime of the particle as standing electromagnetic waves
to the decay products which are electromagnetic waves as well.

The concept of a nuclear cubic lattice provides more than just the
masses of the particles of the $\gamma$-branch.  The masses of the
$\nu$-branch follow from the frequencies in a diatomic neutrino lattice
made of electron and muon neutrinos, with $m(\nu_e) \sim 5$ meV/$c^2$
and $m(\nu_\mu )\sim 50$ meV/$c^2$.  Furthermore, as discussed in [3],
the theory of cubic lattices permits the determination of the potential
of the force that holds the lattice together.  Born and Land\'{e} [17] 
have shown that the potential must have an attractive part over longer
distances and a repulsive part over shorter distances, otherwise the
lattice would not be stable.  Following the steps of Born and Land\'{e} 
we have found in equation (11) of [3] that the attractive and repulsive
terms of the potential in a nuclear lattice differ at the lattice
distance by only $10^{-12}$, if we replace the electric interaction
constant $e^2$ in a crystal by $g^2$ from the weak nuclear force.
Following a paper of Born and Stern [18] we have discussed in [3] also
the force which acts, (in vacuum), between two cubic lattices.  The
attractive forces between two cubic lattices are the sum of all
unsaturated weak forces at the sides of the lattices.  Since there are
about $10^6$ lattice points on a side of the nuclear cubic lattice
considered here, the attractive force of a side of the lattice for one
side of another lattice is $10^6$ times as large as the weak force
acting between two lattice points.  The empirical relation between the
strength of the strong and weak forces is given by the ratio of the
coupling constants, which is $\alpha_s /\alpha_w \approx 10^6$.

Our standing wave model does not only account for the masses of the
mesons and baryons and the antiparticles of the $\gamma$-branch, but
also provides access to an explanation of the weak force which holds the
nuclear lattice together, and the strong force which emanates from the
surface of the particles.  The strong and the weak forces are unified in
this model. \vspace*{1ex}

\noindent \textbf{Acknowledgements.} We are, in particular, grateful to
Professor I.\ Prigogine for his support.  We thank M.~Fink and
L.~Frommhold for discussions.  ELK is grateful to Y.\ Tassoulas for the
calculation of numerous eigenvalues of the elastic oscillations of
cubes.  \vspace{2ex}

\noindent
REFERENCES \\
{\small \mbox{}
[1] L.\ Koschmieder, preceding article. \\ \mbox{}
[2] L.\ Koschmieder, Nuovo Cimento {\bf A99},  555  (1988). \\ \mbox{}
[3] L.\ Koschmieder, Nuovo Cimento {\bf A101},  1017  (1989). \\ \mbox{}
[4] K.\ Wilson,  Phys.\ Rev.\ {\bf D10},  2445  (1974). \\ \mbox{}
[5] D.\ Weingarten, Scient.\ American {\bf 274},  116  (1996). \\ \mbox{}
[6] M.\ Born and Th.\ v.\ Karman,  Phys.\ Zeitschr.\ {\bf 13},  297 
(1912). \\ \mbox{}
[7] M.\ Blackman,  Proc.\ Roy.\ Soc.\ {\bf A148}, 365; 384 (1935). \\ \mbox{}
[8] M.\ Blackman,  in Handbuch der Physik VII/1  (1955) Sec.\ 12. \\ \mbox{}
[9] M.\ Born and K.\ Huang, \textit{Dynamical Theory of Crystal 
Lattices}, (Oxford)  1954. \\ \mbox{}
[10] A.\ Maradudin, E.\ Montroll, G.\ Weiss and I.\ Ipatova,  
\textit{Theory of Lattice Dynamics in the Harmonic Approximation}, 
(Academic Press), 2nd edition,  1971. \\ \mbox{}
[11] J.\ Schwinger,  Phys.\ Rev.\ {\bf 128}, 2425 (1962). \\ \mbox{}
[12] D.\ Perkins,  \textit{Introduction to High-Energy Physics}, 
(Addison Wesley 1982, p.128. \\ \mbox{}
[13] H.\ Frauenfelder and E.\ Henley,  \textit{Subatomic Physics}, 
(Prentice- Hall)  1974, p.\ 128. \\ \mbox{}
[14] A.\ Sommerfeld,  \textit{Vorlesungen \"{u}ber Theoretische Physik}, 
1952, Bd.V, 56. \\ \mbox{}
[15]  P.\ Debye, Ann.\ d.\ Phys.\ {\bf 39},  789  (1912). \\ \mbox{}
[16]  S.\ Bose,  Zeitschr.\ f.\ Phys.\ {\bf 26},  178  (1924). \\ \mbox{}
[17] M.\ Born and A.\ Land\'{e}, Verh.\ Dtsch.\ Phys.\ Ges.\ {\bf 20}, 
210 (1918). \\ \mbox{}
[18] M.\ Born and O.\ Stern, Sitzungsber.\ Preuss.\ Akad.\ Wiss.\ 
{\bf 33}, 901 (1919).}

\end{document}